\documentclass[12pt,a4paper]{amsart}
\usepackage[a4paper,inner=2.8cm,outer=2.5cm,top=2.8cm,bottom=2.5cm]{geometry}                
\usepackage{amsmath,amssymb,amscd,amsthm,amsfonts,anyfontsize,color,dsfont,enumerate,fix-cm,layout,lipsum,lpic,mathrsfs,mdwlist,pigpen,stmaryrd,tensor,tikz,thmtools,xspace,url}
\usepackage{hyperref}
\usepackage[all]{xy}
\usepackage{mathtools}
\usepackage{xfrac}
\theoremstyle{plain}                 
\newtheorem{theorem}{Theorem}[section]

\theoremstyle{definition}

\theoremstyle{remark}

\hypersetup{colorlinks=true,linkcolor=blue,citecolor=purple,filecolor=magenta,urlcolor=cyan}

\def\CP1{\mathbb{C}\mathrm{P}^1}

\allowdisplaybreaks[3]

\newcommand\fixme[1]{\textbf{\color{red} [[ #1 ]]}}

\newcommand\prePositiveCrossing{
  \put(0,0){
    \put(0,0){\qbezier(0,0)(10,10)(20,20)}
    \put(20,0){\qbezier(0,0)(-5,5)(-8,8)}
    \put(0,20){\qbezier(0,0)(5,-5)(8,-8)}
  }
}

\newcommand\preNegativeCrossing{
  \put(0,0){
    \put(0,0){\qbezier(20,0)(10,10)(0,20)}
    \put(0,0){\qbezier(0,0)(5,5)(8,8)}
    \put(20,20){\qbezier(0,0)(-5,-5)(-8,-8)}
  }
}

\newcommand\arrowsUp{
  \put(20,20){\vector(1,1){0}}
  \put(0,20){\vector(-1,1){0}}
}

\newcommand\arrowsRight{
  \put(20,20){\vector(1,1){0}}
  \put(20,0){\vector(1,-1){0}}
}

\newcommand\arrowsLeft{
  \put(0,20){\vector(-1,1){0}}
  \put(0,0){\vector(-1,-1){0}}
}

\newcommand\arrowsDown{
  \put(0,0){\vector(-1,-1){0}}
  \put(20,0){\vector(1,-1){0}}
}

\newcommand\crossingPosUp{
  \prePositiveCrossing
  \arrowsUp
}

\newcommand\crossingNegUp{
  \preNegativeCrossing
  \arrowsUp
}

\newcommand\crossingPosRight{
  \preNegativeCrossing
  \arrowsRight
}

\newcommand\crossingNegDown{
  \preNegativeCrossing
  \arrowsDown
}

\newcommand\crossingPosLeft{
  \preNegativeCrossing
  \arrowsLeft
}

\newcommand\arcUp{
  \put(0,0){\qbezier(0,0)(10,10)(20,0)}
}

\newcommand\arcDown{
  \put(0,0){\qbezier(0,0)(10,-10)(20,0)}
}

\newcommand\arcLeft{
  \put(0,0){\qbezier(0,0)(-10,10)(0,20)}
}

\newcommand\arcRight{
  \put(0,0){\qbezier(0,0)(10,10)(0,20)}
}

\newcommand\arcSmallDown{
  \put(0,0){\qbezier(0,0)(5,-5)(10,0)}
}

\newcommand\arcSmallLeft{
  \put(0,0){\qbezier(0,0)(-5,5)(0,10)}
}

\newcommand\arcSmallRight{
  \put(0,0){\qbezier(0,0)(5,5)(0,10)}
}

\newcommand\arcMedUp{
  \put(0,0){\qbezier(0,0)(15,15)(30,0)}
}

\newcommand\slashPos{
  \put(0,0){\qbezier(0,0)(10,10)(20,20)}
}

\newcommand\slashNeg{
  \put(-20,0){\qbezier(20,0)(10,10)(0,20)}
}

\newcommand\bendDownRight{
  \put(0,0){\qbezier(0,0)(0,-10)(10,-20)}
}

\newcommand\bendDownLeft{
  \put(0,0){\qbezier(0,0)(0,-10)(-10,-20)}
}

\newcommand\bendUpRight{
  \put(0,0){\qbezier(0,0)(0,10)(10,20)}
}

\newcommand\bendUpLeft{
  \put(0,0){\qbezier(0,0)(0,10)(-10,20)}
}

\begin{document}

\title{Evolution for Khovanov polynomials for figure-eight-like family of knots}
\author{Petr Dunin-Barkowski}
\address{Higher School of Economics, Moscow, Russia}
\email{ptdbar@gmail.com}
\author{Aleksandr Popolitov}
\address{Department of Physics and Astronomy, Uppsala University,\\
Box 516, SE-75120 Uppsala, Sweden.}
\email{popolit@gmail.com}
\author{Svetlana Popolitova}
\address{Sub-department of Financial management, MSTU ``STANKIN'', Moscow, Russia}
\email{spopolitova@yandex.ru}

{\begin{abstract}
    %% \fixme{Summarize what we do in this paper}
    We look at how evolution method deforms, when one considers Khovanov polynomials
    instead of Jones polynomials. We do this for the figure-eight-like knots
    (also known as 'double braid' knots, see arXiv:1306.3197)
    -- a two-parametric family of knots which ``grows'' from the figure-eight knot
    and contains both two-strand torus knots and twist knots.
    We prove that parameter space splits into four chambers, each with its own evolution,
    and two isolated points.
    Remarkably, the evolution in the Khovanov case features an extra eigenvalue,
    which drops out in the Jones $(t \rightarrow -1)$ limit.
\end{abstract}}

\today

\maketitle

\flushbottom

{\section{Introduction}
  {One of the central questions in knot theory is whether two knots are topologically equivalent.
    Currently, the most effective way to
    determine this is to compute some polynomial knot invariant for each knot:
    if the invariants differ, knots are for sure not topologically equivalent.
    Complementary statement is not true: if invariants are the same it does not mean
    that the knots are equivalent. So, in knot theory one seeks stronger and stronger
    knot invariants that have the power to distinguish more and more knots.
  }
  
  {Perhaps, the most widely known polynomial knot invariants (knot polynomials) are
    the so-called Jones \cite{JonesPolynomialDef} and HOMFLY \cite{HOMFLYPolynomialDef,PTPolynomialDef} polynomials.
    One of the ways to compute (and define) them is quite elementary
    -- one just repeatedly applies skein relations
    
    %% \begin{verbatim}
    %% picture of the skein relations for Jones and HOMFLY
    %% \end{verbatim}
    \begin{align} \label{eq:skein-relation}
    \begin{picture}(300,40)(-50,-10)
      \put(0,0){\put(-12,6){$A$} \thicklines \crossingPosUp}
      \put(40,7){$-$}
      \put(80,0){\put(-18,6){$A^{-1}$} \thicklines \crossingNegUp}
      \put(120,7){$=$}
      \put(190,0){\put(-48,6){$\left(q - q^{-1}\right)$} \thicklines
        \arcRight \put(0,20){\vector(-1,1){0}}
        \put(20,0){\arcLeft} \put(20,20){\vector(1,1){0}}
      }
    \end{picture}
    \end{align}

    \noindent and Reidemeister moves to express knot polynomial for a given planar diagram
    through polynomials for simpler diagrams.
    %% can be computed via skein relations, at least in principle.
  }

  {
    {
      %% For usual Jones, as well as HOMFLY, polynomials, there is an evolution method
      A surprising and indirect consequence of the above naive definition is the existence
      of the so-called \textit{evolution method} for both Jones and HOMFLY polynomials
      (see, for instance, \cite[Introduction]{AnoM}; this story is also well-explained in \cite{MMM-evo-diff-colored}).
      This is the statement that, whenever
      an $m$-strand braid is present somewhere in the planar diagram $\mathcal{K}$ of some knot,
      the knot polynomial $P$ (either Jones or HOMFLY)  depends on the number $n$ of windings of the braid
      in quite a simple way

      %% \begin{verbatim}
      %% the schematic formula for the evolution method
      %% \end{verbatim}
      \begin{align} \label{eq:evolution-method}
        P^{\mathcal{K}}(n) = \sum_{\lambda \vdash m} C^{\mathcal{K}}(\lambda) \left((\pm)_\lambda q^{\text{power}(\lambda)}\right)^n
      \end{align}

      \noindent Here sum is over partitions $\lambda$ of the number of strands $m$ (i.e. over Young diagrams),
      and sign~'$(\pm)_\lambda$'
      and the exponent~'$\text{power}(\lambda)$' of an eigenvalue are universal; they depend only on partition~$\lambda$
      but not on the diagram $\mathcal{K}$.
      In fact, they are simple
      combinatorial expressions of the shape of~$\lambda$.
      The \textit{evolution coefficients}~$C^{\mathcal{K}}(\lambda)$, on the contrary,
      depend both on the knot diagram~$\mathcal{K}$ and the partition~$\lambda$.
    }

    {%% Series for positive and negative crossings merge into one integer series.
      %% This is a powerful hint that the RT formalism exists
      A subtle feature of the above formula is that number of windings $n$ can be
      arbitrary integer -- the evolution coefficients $C^{\mathcal{K}}(\lambda)$
      \textbf{do not change} as one goes from negative number of crossings
      (i.e. crossings in a different direction)
      to zero crossings to positive number of crossings
      in the braid. The formula seamlessly interpolates between these three cases.
      This feature is, in fact, a manifestation of another, more conceptual and deep,
      definition of Jones and HOMFLY polynomials
      -- through the so-called Reshetikhin-Turaev (RT) formalism \cite{RTpaper}.
    }
    
    { %% The RT-formalism in a nutshell
      In the RT-formalism one associates a certain linear operator,
      the so-called $\mathcal{R}$-matrix,
      to each positive crossing
      and to each negative crossing its inverse. The factors $\left((\pm)_\lambda q^{\text{power}(\lambda)}\right)$
      in the formula \eqref{eq:evolution-method}
      are then naturally reinterpreted as eigenvalues of the $\mathcal{R}$-matrix in the irreducible representation
      corresponding to the Young diagram $\lambda$. The knot polynomial itself
      is just a tensor contraction of several $\mathcal{R}$-matrices,
      in the order dictated by the planar diagram $\mathcal{K}$.
    }

    { RT-formalism allows one to establish a link between knot theory and physics:
      knot polynomials turn out to be Wilsonian averages in Chern-Simons theory
      (see, for instance, \cite{Ano-r-matrix-approaches} for a review).
      This is most easily seen in the so-called \textit{temporal gauge} \cite{SmirnovMorozovTempGauge}.
      The gauge theory point of view immediately leads to the generalization
      of both Jones and HOMFLY polynomials to the \textit{colored} case, where one decorates the knot with some
      representation of the gauge group. The skein relation description is not available in the colored case
      and from the naive skein relations \eqref{eq:skein-relation} it is not at all easy to guess, that
      the colored generalization should even exist.
    }
  }

  {
    {The (uncolored, or fundamental)
      Jones and HOMFLY polynomials have a \textit{homological} generalization -- Khovanov \cite{KhovanovDefinition}
      and Khovanov-Rozansky \cite{KhovanovRozanskyDefinition1,KhovanovRozanskyDefinition2}
      polynomials, respectively. The definition of these polynomials (especially the Khovanov-Rozansky one)
      seems much more elaborate than definition of their non-homological analogs -- one needs to calculate
      the homology groups of
      a certain differential complex, built out of the planar diagram $\mathcal{K}$.
      While in the case of Khovanov polynomial at least the construction of the complex is well-understood
      \cite{BarNatan},
      in the Khovanov-Rozansky case even explicit construction of the linear spaces in the complex
      presents difficulties (each space is a factor of an infinite-dimensional space of so-called foams
      over infinitely many relations \cite{RobertWagnerSymmetricKhR}).
      Explicit construction of maps between the spaces is even more difficult.
      Hence, one is tempted to search for alternative definitions for Khovanov and Khovanov-Rozansky
      polynomials \cite{DolMorI,DolMorII,DolMorIII,AnoMorTowardsRMatKhov,AnoSolCount},
      partly motivated by the empirical observation that \textit{answers}
      for both Khovanov and Khovanov-Rozansky polynomials (in known examples)
      seem much simpler than their cumbersome definitions would lead one to expect.
    }

    {A natural question on this path is
      whether the relevant generalization of the RT formalism
      exists for the Khovanov and Khovanov-Rozansky polynomials.
      This homological RT formalism is believed to be related
      to the so-called \textit{refined} \cite{ArthShakDAHA} Chern-Simons theory,
      which is at the moment defined only for the simplest examples of knots.
    }
  }
  
  { %% The evolution for Kh and KhR
    {Even more naively, one may wonder, whether something like evolution formula \eqref{eq:evolution-method}
      exists for Khovanov and Khovanov-Rozansky polynomials.
      In \cite{AnoM} this was investigated for the simplest case of torus knots.
      Surprisingly, it was observed that the symmetry between positive and negative crossings is broken
      -- an essential difference with the Jones and HOMFLY case.
      Moreover, from studies of superpolynomials (the $N \rightarrow \infty$ stable component of Khovanov-Rozansky polynomials)
      for torus knots (see \cite{DBMMSS-superpolys-toric})
      one knows that for them the symmetry between positive and negative crossings is \textit{not} broken,
      which makes the observed breaking at finite $N$ even more unexpected.
    }
  }

  { %% ### Introduction : results
    {In this short note we look at what happens for a slightly
      more complicated family of knots (see Section~\ref{sec:fig-eight-like-knots}).
      This family is inspired by the figure 8 knot (the first non-torus knot)
      and one is allowed to insert arbitrary two-strand braids in place of its upper and lower pairs of crossings.}

    {We concentrate only on Khovanov polynomials and completely ignore Khovanov-Rozansky ones.
      First of all, Khovanov polynomials are easier to calculate:
      there is a computer program readily available in the ``KnotTheory'' package for Wolfram Mathematica,
      which is available on Katlas~\cite{katlas}, though there are some caveats (see Section~\ref{sec:caveats}).
      Second of all, non-trivial new phenomena are seen already in Khovanov case, so this generalization
      is sufficiently interesting.
    }

    {
      {We find the following picture (see Section~\ref{sec:phase-diagram}, Theorem~\ref{thm:evolution-method}).
        For some suitable subfamily (see Section~\ref{sec:fig-eight-like-knots}) we observe that:
        \begin{itemize}
        \item evolution is preserved in the \textit{chambers} on the parameter plane;
          there are also 2 isolated points, where the knot becomes two unlinked unknots,
        \item if one interprets the formulas as coming from some sort of RT-formalism,
          one concludes that \textit{fundamental} $\mathcal{R}$-matrix must have one more eigenvalue:
          $(-1) t q^3$, in addition to $t q^3$ and $q$. The coefficient in front of this
          eigenvalue is always proportional to $(t+1)$ and vanishes when one goes to
          the Jones limit $t \rightarrow -1$,
        \item transitions between chambers are tricky: some coefficients in front of eigenvalues
          just get multiplied by some constants, while others re-glue in a more elaborate way.
        \end{itemize}
      }
    }

    { We prove our evolution formulas (Theorem~\ref{thm:evolution-method}) in Section~\ref{sec:proof}. The proof relies
      on the fact that all knots and links in the family we consider are alternating, and on the reconstruction
      theorem \cite[Theorem 4.5]{Lee-endomorphism} that allows one to completely determine Khovanov polynomial
      from the Jones polynomial for the alternating knot/link, provided the signature is known.
      The symmetry breaking in the evolution method is ultimately traced to the jumps in the signature.
    }

    { Forward references throughout the introduction provide sufficient description
      of the organization of the paper, so ``this paper is organized as follows'' paragraph
      is really unnecessary.
    }
    
    %% {We then, in Section~\ref{sec:comparison-with-bngkh-conjectures}, consider,
    %%   how our observed (conjectural) evolution formulas compare with the BNGKh-conjecture \cite[Conjecture 1,2]{BarNatan}.
    %%   These conjectures imply that Khovanov polynomial for any prime alternating knot is completely determined
    %%   by its Jones polynomial. Since all knots in our two-parametric family are prime and alternating,
    %%   and evolution works for entire two-parametric family for the Jones polynomials, it is very interesting
    %%   to see, where exactly does it break down when we go from Jones level to Khovanov level.
    %%   We find good agreement of our conjecture and BNGKh-conjecture.
    %% }
    
    %% {This note is experimental mathematics. We report phenomena we observe with help of computer experiments,
    %%   but at the moment we don't yet have any explanation (the homological RT-formalism)
    %%   of \textit{why} the things are the way they are.
    %%   By writing this note we hope to bring more research attention to this exciting topic.
    %% }
  }

  {\subsection*{Acknowledgements} We would like to thank A.Anokhina, A.Morozov, Y.Zenkevich
    and M.Khovanov
    for stimulating discussions.
    The research of A.P. is supported in part by
    Vetenskapsr\r{a}det under grant \#2014-5517,
    by the STINT grant,
    by the grant  ``Geometry and Physics"  from the Knut and Alice Wallenberg foundation
    and by RFBR grants 16-01-00291 and 18-31-20046 mol\_a\_ved.
    The research of P.D.-B. is supported in part by RFBR grants 18-01-00461 A and 18-31-20046 mol\_a\_ved.
  }
}

%% ### We don't really need this section -- we just cite the AnoM everywhere
%% {\section{Evolution method} \label{sec:evolution-method}
%%     \fixme{The summary of how the evolution method works for usual HOMFLY polynomials}
%% }

%% ### Test of the various crossings
%% \begin{picture}(300,75)(0,0)
%%   \thicklines
%%   \put(100,0){
%%     \put(0,0){\positiveCrossingUp}
%%     \put(30,0){\negativeCrossingUp}}
%%   \put(160,0){
%%     \put(0,0){\positiveCrossingRight}
%%     \put(30,0){\negativeCrossingRight}}
%%   \put(220,0){
%%     \put(0,0){\positiveCrossingDown}
%%     \put(30,0){\negativeCrossingDown}}
%%   \put(280,0){
%%     \put(0,0){\positiveCrossingLeft}
%%     \put(30,0){\negativeCrossingLeft}}
%%   \put(350,0){\arcUp \arcDown \arcLeft \arcRight}
%% 
%% \end{picture}

\newcommand\figureEightLikeTemplate{
  { %% ### Arcs in the middle of the knot
    \put(0,20){\arcSmallLeft}
    \put(60,20){\arcSmallRight}
    \put(0,30){\slashPos}
    \put(60,30){\slashNeg}
  }
  { %% ### The left big arc
    \put(-10,0){\arcSmallDown}
    \put(-20,20){\bendDownRight}
    \put(-20,20){\line(0,1){70}}
    \put(-20,90){\bendUpRight}
    \put(-10,110){\arcMedUp}
  }
  { %% ### The right big arc
    \put(60,0){\arcSmallDown}
    \put(80,20){\bendDownLeft}
    \put(80,20){\line(0,1){70}}
    \put(80,90){\bendUpLeft}
    \put(40,110){\arcMedUp}
  }
}

\newcommand\theBottomBraid{
  { %% ### The bottom braid
    \linethickness{0.5mm}
    \put(0,0){\line(0,1){20}}
    \put(0,0){\line(1,0){60}}
    \put(0,20){\line(1,0){60}}
    \put(60,0){\line(0,1){20}}
    \put(27,6){$b$}
  }
}

\newcommand\theTopBraid{
  { %% ### The top braid
    \linethickness{0.5mm}
    \put(0,0){\line(0,1){60}}
    \put(0,0){\line(1,0){20}}
    \put(0,60){\line(1,0){20}}
    \put(20,0){\line(0,1){60}}
    \put(7,26){$a$}
  }
}

{\section{The figure-eight-like knots} \label{sec:fig-eight-like-knots}
  {We consider the following family of planar diagrams, inspired by the figure-8 knot

    %% \begin{verbatim}
    %% picture of the figure eight going to the family
    %% \end{verbatim}
    
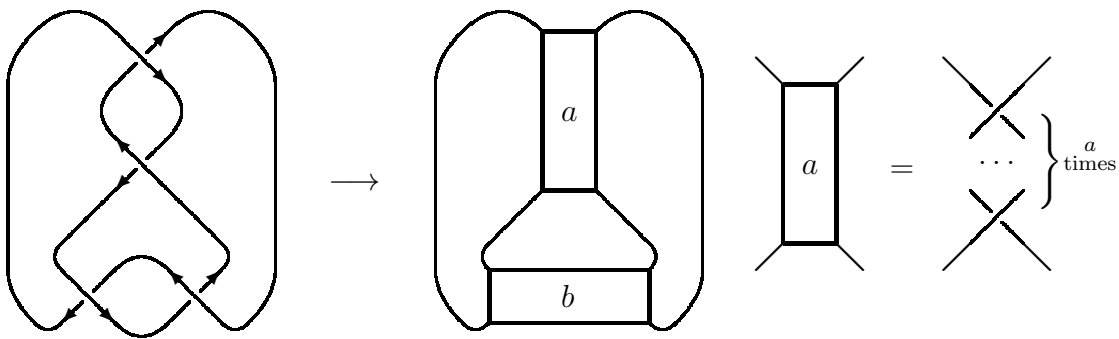
\begin{picture}(300,130)(-20,-5)
      %% ### Alright, let's start to draw a figure 8
      \put(0,0){ %% ### A copy of 'vanilla' figure 8
        \thicklines
        \figureEightLikeTemplate

        \put(0,0){
          \put(0,0){\crossingNegDown}
          \put(20,0){\arcDown}
          \put(20,20){\arcUp}
          \put(40,0){\crossingNegUp}
        }
        \put(20,50){
          \put(0,0){\crossingPosLeft}
          \put(0,40){\crossingPosRight}
          \put(0,20){\arcLeft}
          \put(20,20){\arcRight}
        }
      }

      \put(100,50){$\longrightarrow$}
      
      \put(160,0){ %% ### figure 8 with braids thrown in
        \thicklines
        \figureEightLikeTemplate

        \theBottomBraid
        \put(20,50){\theTopBraid}
      }

      \put(250,0){ %% ### The standalone A-braid
        \put(20,30){
          \theTopBraid
          \thicklines
          \put(0,0){\line(-1,-1){10}}
          \put(20,0){\line(1,-1){10}}
          \put(0,60){\line(-1,1){10}}
          \put(20,60){\line(1,1){10}}
        }
      }
      \put(310,55){$=$}
      \put(320,0){ %% ### The explained A-braid
        \put(20,30){
          \thicklines
          \put(0,0){\line(-1,-1){10}}
          \put(20,0){\line(1,-1){10}}
          \put(0,60){\line(-1,1){10}}
          \put(20,60){\line(1,1){10}}
          \put(0,0){\prePositiveCrossing}
          \put(0,40){\prePositiveCrossing}
          \put(3,30){$\dots$}

          \put(25,28){$\Bigg\}$}
          \put(35,30){$\substack{a \\ \text{times}}$}
        }
      }
    \end{picture}
    
\noindent i.e. we allow arbitrary 2-strand braid to be inserted in place of higher or lower pair of crossings
of the figure eight knot. This family was also considered in \cite{MMM-evo-diff-colored}, where it was called
``double braid'' family of knots, and where in particular a conjecture was made about explicit form of
colored HOMFLY polynomials for this family in any symmetric representation $[r]$.
  }

  {This family is interesting to consider, since it includes both torus and non-torus knots
    (so we will not observe something that is specific just to torus case).
    It is also two-parametric, which allows us to see, how evolutions w.r.t the
    two braid-winding parameters $a$ and $b$ interplay.
    Moreover, it was instrumental in finding an explicit relation between so-called
    inclusive and exclusive Racah matrices (see \cite{MMMS-racah-hidden-integrability})
  }

  {Depending on the parity of the braid parameters~$a$ and $b$, strands are oriented differently
    (so $a$- and $b$- braids can be both parallel and antiparallel).
    Moreover, in case both $a$ and $b$ are odd the diagram is a link, so orientations of its two
    components can be chosen independently.
    %% ### WONTDO -- picture of a particular choice of orientation is enough
    %% \begin{verbatim}
    %% picture of the different possible choices of orientation
    %% \end{verbatim}
  }

  {In what follows we choose a particular orientation of strands
    
    
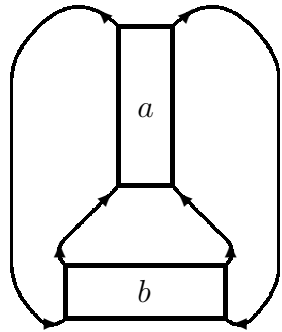
\begin{picture}(300,130)(-20,-5)
      \put(160,0){
        \thicklines
        \figureEightLikeTemplate

        \theBottomBraid
        \put(20,50){\theTopBraid}

        { %% ### Orientations for the bottom braid
          \put(60,0){\put(2,-2){\vector(-1,0){0}}}
          \put(0,20){\put(-2,9){\vector(0,1){0}}}
          \put(0,0){\put(-2,-2){\vector(1,0){0}}}
          \put(60,20){\put(2,9){\vector(0,1){0}}}
        }
        { %% ### Orientations for the top braid
          \put(20,50){\put(-2,-2){\vector(1,1){0}}}
          \put(40,50){\put(2,-2){\vector(-1,1){0}}}
          \put(20,110){\put(-8,7){\vector(-1,1){0}}}
          \put(40,110){\put(8,7){\vector(1,1){0}}}
        }
      }
    \end{picture}
          
    %% concentrate on the sub-family \fixme{$a$-arbitrary and $b$-odd + this(?) choice of orientation}:
    \noindent With this choice $a$ can be arbitrary integer, while $b$ can only be odd.
    On the example of the $a$-braid we are able to see what happens when we add a single crossing to the diagram.
    If the representation theory is at all applicable to Khovanov polynomials,
    then in some sense we must have, that in the $a$-braid there is the representation $[1]\otimes[1] = [1,1] \oplus [2]$ running,
    while in the $b$-braid there is the representation $[1]\otimes\overline{[1]} = \emptyset \oplus \text{\textit{adjoint}}$ running.
    Of course, it would be interesting in future to consider a family of knots that allows
    to add single crossings to more that one of its braids and see whether this produces some interesting effects.
}

  {\section{The phase diagram} \label{sec:phase-diagram}
  {We find (see Section~\ref{sec:proof} for a proof) that the space of parameters splits into 4 regions and 2 isolated points
    
    %% \begin{verbatim}
    %% picture of the phase diagram
    %% \end{verbatim}
    
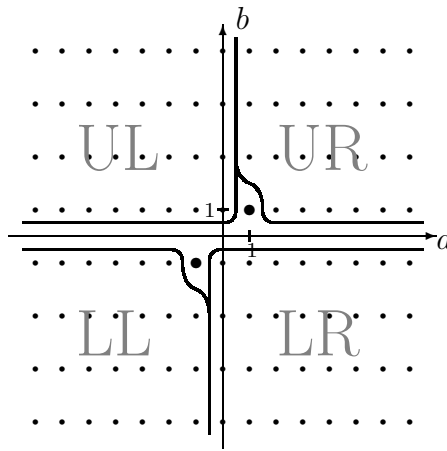
\begin{picture}(300,170)(0,-10)
      \put(200,75){
        {%% ### coordinate axes
          \put(-80,0){\vector(1,0){160}}
          \put(0,-80){\vector(0,1){160}}
          \put(80,-5){$a$}
          \put(5,78){$b$}
        }
        { %% ### The scale on the coordinate axes
          \put(10,-2){\line(0,1){4}}
          \put(-2,10){\line(1,0){4}}
          \put(9,-6){${}_1$}
          \put(-7,9){${}_1$}
        }
        { %% ### Two isolated special points
          \put(10,10){\circle*{4}}
          \put(-10,-10){\circle*{4}}
        }
        { %% ### the grid of integer points
          %% \multiput(10,10)(10,0){7}{\circle*{1}}
          \multiput(-70,10)(10,0){15}{\circle*{2}}
          \multiput(-70,30)(10,0){15}{\circle*{2}}
          \multiput(-70,50)(10,0){15}{\circle*{2}}
          \multiput(-70,70)(10,0){15}{\circle*{2}}
          \multiput(-70,-10)(10,0){15}{\circle*{2}}
          \multiput(-70,-30)(10,0){15}{\circle*{2}}
          \multiput(-70,-50)(10,0){15}{\circle*{2}}
          \multiput(-70,-70)(10,0){15}{\circle*{2}}
        }
        { %% ### The upper-left region
          \thicklines
          \put(-75,5){\line(1,0){75}}
          \put(0,0){\qbezier(0,5)(5,5)(5,10)}
          \put(5,10){\line(0,1){65}}
        }
        { %% ### The upper-right region
          \thicklines
          \put(20,10){\qbezier(-5,0)(-5,-5)(0,-5)}
          \put(20,5){\line(1,0){55}}
          \put(10,30){\qbezier(-5,0)(-5,-8)(0,-10)}
          \put(10,10){\qbezier(5,0)(5,8)(0,10)}
        }
        { %% ### The lower-right region
          \thicklines
          \put(0,-5){\line(1,0){75}}
          \put(0,-10){\qbezier(0,5)(-5,5)(-5,0)}
          \put(-5,-10){\line(0,-1){65}}
        }
        { %% ### The lower-right region
          \thicklines
          \put(-20,-10){\qbezier(5,0)(5,5)(0,5)}
          \put(-20,-5){\line(-1,0){55}}
          \put(-10,-30){\qbezier(5,0)(5,8)(0,10)}
          \put(-10,-10){\qbezier(-5,0)(-5,-8)(0,-10)}
        }
        \put(-5,-10){ %% ### Labels of the regions
          \put(-50,35){\color{gray}{\Huge{UL}}}
          \put(25,35){\color{gray}{\Huge{UR}}}
          \put(25,-35){\color{gray}{\Huge{LR}}}
          \put(-50,-35){\color{gray}{\Huge{LL}}}
        }
      }
    \end{picture}
  
    \noindent The isolated points are $(1,1)$ and $(-1,-1)$ and in what follows we will refer to the
    four regions as UL, UR, LL and LR for upper-left, upper-right, lower-left and lower-right, respectively.
  }

  {Dependence of Khovanov polynomial on the number of braid windings $a$ and $b$ is quite remarkable:
    \begin{theorem}\label{thm:evolution-method}
    In each of the regions UL, UR, LL and LR dependence of the Khovanov polynomial on the braid parameters $a$ and $b$
    is consistent with the evolution method
    %% \begin{verbatim}
    %% formula with the 6-element matrix
    %% \end{verbatim}
    \begin{align} \label{eq:evolution-formula}
      Kh_{\substack{a \hfill \\ b=2 k+1}} =
      \left(\begin{array}{cc} (1)^b & (t q^2)^b \end{array}\right)
      \left(\begin{array}{ccc} M_{1,1} & M_{1,2} & M_{1,3} \\ M_{2,1} & M_{2,2} & M_{2,3} \end{array}\right)
      \left(\begin{array}{c} q^{-a} \\ (t q^3)^{-a} \\ (-t q^3)^{-a} \end{array}\right)
    \end{align}

    \noindent but the matrices of coefficients $M$ are different among the four regions
    %% \begin{verbatim}
    %% explicit formulas for the matrices
    %% \end{verbatim}
    
    \begin{align} \label{eq:evolution-matrices}
      M_{UL} & = \left(\begin{array}{ccc}
        \frac{q^8 t^3+q^6 t^3-q^4 t^2-q^4 t+q^2+1}{\left(q^2 t-1\right)^2 \left(q^3 t+q\right)}
        & \frac{- q^5 t^2-q t}{\left(q^2 t-1\right)^2 \left(q^2 t+1\right)}
        & 0
        \\ \frac{-q^8 t^3+q^6 t^2-q^4 t^2-q^4 t+q^2 t-1}{\left(q^2 t-1\right)^2 \left(q^3 t+q\right)}
        & \frac{2 q^8 t^3+q^6 t^3-q^6 t^2-q^2 t+q^2+2}{2 \left(q^2 t-1\right)^2 \left(q^3 t+q\right)}
        & \frac{- q t-q}{2 q^2 t+2}
      \end{array}\right) \\ 
      M_{UR} & = \left(\begin{array}{ccc}
        \frac{q^8 t^4-2 q^6 t^3-q^6 t^2+q^2 t-1}{q t \left(q^2 t-1\right)^2 \left(q^2 t+1\right)}
        & \frac{q^5 t+q}{\left(q^2 t-1\right)^2 \left(q^2 t+1\right)}
        & 0
        \\ \frac{q^8 t^3-q^6 t^2+q^4 t^2+q^4 t-q^2 t+1}{q t \left(q^2 t-1\right)^2 \left(q^2 t+1\right)}
        & \frac{q^8 t^4-q^8 t^3-2 q^6 t^3-q^4 t^2-q^4 t+2 q^2 t-2}{2 q t \left(q^2 t-1\right)^2 \left(q^2 t+1\right)}
        & \frac{-q^3 t-q^3}{2 q^2 t+2}
      \end{array}\right) \\ \notag
      M_{LL} & = \left(\begin{array}{ccc}
        \frac{-q^8 t^4+q^6 t^3-q^2 t^2-2 q^2 t+1}{\left(q^2 t-1\right)^2 \left(q^3 t+q\right)}
        & \frac{q^5 t^3+q t^2}{\left(q^2 t-1\right)^2 \left(q^2 t+1\right)}
        & 0
        \\ \frac{q^8 t^4-q^6 t^3+q^4 t^3+q^4 t^2-q^2 t^2+t}{\left(q^2 t-1\right)^2 \left(q^3 t+q\right)}
        & \frac{-2 q^8 t^4+2 q^6 t^3-q^4 t^3-q^4 t^2-2 q^2 t-t+1}{2 \left(q^2 t-1\right)^2 \left(q^3 t+q\right)}
        & \frac{-t-1}{2 \left(q^3 t+q\right)}
      \end{array}\right) \\ \notag
      M_{LR} & = \left(\begin{array}{ccc}
        \frac{q^8 t^3+q^6 t^3-q^4 t^2-q^4 t+q^2+1}{\left(q^2 t-1\right)^2 \left(q^3 t+q\right)}
        & \frac{-q^5 t^2-q t}{\left(q^2 t-1\right)^2 \left(q^2 t+1\right)}
        & 0
        \\ \frac{-q^8 t^3+q^6 t^2-q^4 t^2-q^4 t+q^2 t-1}{\left(q^2 t-1\right)^2 \left(q^3 t+q\right)}
        & \frac{2 q^8 t^3+q^6 t^3-q^6 t^2-q^2 t+q^2+2}{2 \left(q^2 t-1\right)^2 \left(q^3 t+q\right)}
        & \frac{-q t-q}{2 q^2 t+2}
      \end{array}\right)
    \end{align}
    \end{theorem}

    We give a proof of this theorem in Section~\ref{sec:proof}.
  }

  { %% ### The extra eigenvalue
    {The fact that the space of parameters splits into chambers with phase transitions between them
      is striking by itself, however, evolution formula \eqref{eq:evolution-formula} has
      another notable feature: there are \textit{three} eigenvalues for the evolution w.r.t $a$-parameter, whereas
      in the HOMFLY (and Jones) case there were only two. What is the meaning of the third eigenvalue
      in the relevant $t$-deformation of the representation theory
      (where, naively, from the decomposition $[1]\otimes[1] = [2]\oplus[1,1]$ one expects \textit{two} eigenvalues)
      is an interesting question for future research.
    }

    {The coefficient in front of extra eigenvalue turns out to be always proportional
      to $(t+1)$, so it never contributes to the Jones ($t \rightarrow -1$) limit.
    }
  }

  { %% ### How do matrix elements transform between chambers
    {One immediately sees that some matrix elements in \eqref{eq:evolution-matrices} transform
      quite simply when one goes between the chambers -- they are just multiplied by some scalars.
      Namely, one has the following relations
      \begin{align}
        M_{UR\ 1,2} & = (-t^{-1}) M_{UL\ 1,2}; \ \ M_{UL\ 1,2} = M_{LR\ 1,2} = (-t^{-1}) M_{LL\ 1,2}
        \\ \notag
        M_{UR\ 2,1} & = (-t^{-1}) M_{UL\ 2,1}; \ \ M_{UL\ 2,1} = M_{LR\ 2,1} = (-t^{-1}) M_{LL\ 2,1}
        \\ \notag
        M_{UR\ 2,3} & = q^2 M_{UL\ 2,3}; \ \ \ \ \ \ \ \ M_{UL\ 2,3} = M_{LR\ 2,3} = q^2 M_{LL\ 2,3}
      \end{align}
      \noindent Note that the scalar factor is the same when going from LL-chamber to LR or UL
      and when going from LR or UL to UR. At the moment we don't know why is this so and whether it is always so.
    }

    {On the other hand, coefficients $M_{1,1}$ and $M_{2,2}$ transform in a more complicated way.
      If one plots the (Laurent) polynomials that result from multiplication of the $M_{1,1}$ matrix elements
      by $\left(q^2 t-1\right)^2 \left(q^3 t+q\right)$ on the $(q,t)$ Newton plane one sees the following
      picture (when two points are drawn close to the same integer point that means that they are both
      at this integer point, just are drawn this way so as not to clash)
      
      
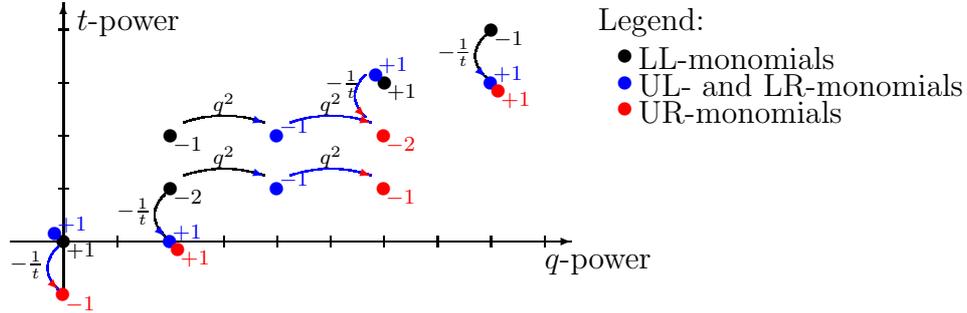
\begin{picture}(300,130)(-60,-30)
        \def\ptDiam{5}
        { %% ### coordinate axes
          \put(-20,0){\vector(1,0){210}}
          \put(0,-20){\vector(0,1){110}}
          \multiput(20,0)(20,0){9}{\put(0,-2){\line(0,1){4}}}
          \multiput(0,20)(0,20){4}{\put(-2,0){\line(1,0){4}}}
          \put(180,0){\put(0,-10){$q$-power}}
          \put(0,80){\put(5,0){$t$-power}}
        }
        { %% ### The legend
          \put(200,80){\put(0,0){Legend:}
            \put(10,0){
              \put(0,-10){{\color{black}\circle*{\ptDiam}} \put(0,-5){LL-monomials}}
              \put(0,-20){{\color{blue}\circle*{\ptDiam}} \put(0,-5){UL- and LR-monomials}}
              \put(0,-30){{\color{red}\circle*{\ptDiam}} \put(0,-5){UR-monomials}}
            }
          }
        }
        \put(0,0){ %% ### LL-points
          \def\delX{-4}
          \def\delY{-4}
          \put(0,0){\circle*{\ptDiam} \put(\delX,\delY){${}_{+1}$}}
          \put(40,20){\circle*{\ptDiam} \put(\delX,\delY){${}_{-2}$}}
          \put(40,40){\circle*{\ptDiam} \put(\delX,\delY){${}_{-1}$}}
          \put(120,60){\circle*{\ptDiam} \put(\delX,\delY){${}_{+1}$}}
          \put(160,80){\circle*{\ptDiam} \put(\delX,\delY){${}_{-1}$}}
        }
        \put(0,0){ %% ### LR- or UL-points
          \def\delX{-4}
          \def\delY{3}
          \color{blue}
          \put(-12,0){ %% ### Correcting shift that is needed to some reason...
            \put(-3,3){\circle*{\ptDiam} \put(\delX,\delY){${}_{+1}$}}
            \put(40,0){\circle*{\ptDiam} \put(\delX,\delY){${}_{+1}$}}
            \put(80,20){\circle*{\ptDiam} \put(\delX,\delY){${}_{-1}$}}
            \put(80,40){\circle*{\ptDiam} \put(\delX,\delY){${}_{-1}$}}
            \put(120,60){\put(-3,3){\circle*{\ptDiam} \put(\delX,\delY){${}_{+1}$}}}
            \put(160,60){\put(0,0){\circle*{\ptDiam} \put(\delX,\delY){${}_{+1}$}}}
          }
        }
        \put(0,0){ %% ### UR-points
          \def\delX{-4}
          \def\delY{-4}
          \color{red}
          \put(-12,0){ %% ### Correcting shift that is needed to some reason...
            \put(0,-20){\circle*{\ptDiam} \put(\delX,\delY){${}_{-1}$}}
            \put(40,0){\put(3,-3){\circle*{\ptDiam} \put(\delX,\delY){${}_{+1}$}}}
            \put(120,20){\circle*{\ptDiam} \put(\delX,\delY){${}_{-1}$}}
            \put(120,40){\circle*{\ptDiam} \put(\delX,\delY){${}_{-2}$}}
            \put(160,60){\put(3,-3){\circle*{\ptDiam} \put(\delX,\delY){${}_{+1}$}}}
          }
        }
        \put(0,0){ %% ### the arrows of coefficient flows
          { %% ### the q^2-group
            \put(40,40){
              \qbezier(5,5)(20,10)(35,5)
              \put(35,5){\color{blue}\vector(3,-1){0}}
              \put(16,5){${}^{q^2}$}
            }
            \put(40,20){
              \qbezier(5,5)(20,10)(35,5)
              \put(35,5){\color{blue}\vector(3,-1){0}}
              \put(16,5){${}^{q^2}$}
            }
            \put(80,40){
              \put(0,0){\color{blue}\qbezier(5,5)(20,10)(35,5)}
              \put(35,5){\color{red}\vector(3,-1){0}}
              \put(16,5){${}^{q^2}$}
            }
            \put(80,20){
              \put(0,0){\color{blue}\qbezier(5,5)(20,10)(35,5)}
              \put(35,5){\color{red}\vector(3,-1){0}}
              \put(16,5){${}^{q^2}$}
            }
          }
          { %% ### the (-t^{-1})-group
            \put(40,20){
              \qbezier(-2,-2)(-10,-10)(-2,-18)
              \put(-2,-18){\color{blue}\vector(1,-1){0}}
              \put(-20,-15){${}^{-\frac{1}{t}}$}
            }
            \put(0,0){
              \put(0,0){\color{blue} \qbezier(-2,-2)(-10,-10)(-2,-18)}
              \put(-2,-18){\color{red}\vector(1,-1){0}}
              \put(-20,-15){${}^{-\frac{1}{t}}$}
            }
            \put(160,80){
              \put(0,0){\color{black} \qbezier(-2,-2)(-10,-10)(-2,-18)}
              \put(-2,-18){\color{blue}\vector(1,-1){0}}
              \put(-20,-15){${}^{-\frac{1}{t}}$}
            }
            \put(120,60){
              \put(-5,5){
                \put(0,0){\color{blue} \qbezier(-2,-2)(-10,-10)(-2,-18)}
                \put(-2,-18){\color{red}\vector(1,-1){0}}
                \put(-17,-12){${}^{-\frac{1}{t}}$}
              }
            }
          }
        }
      \end{picture}

      \noindent One of the possible ways of how different monomials ``flow'' when one goes from LL-chamber
      to LR-chamber to UR-chamber is denoted with arrows. It's interesting that this flow features multiplication
      by same coefficients $q^2$ and $-t^{-1}$ that appear in phase transition of the less complicated matrix elements
      $M_{1,2}$, $M_{2,1}$ and $M_{1,3}$. The picture of the Newton plane for the elements $M_{2,2}$ is similar.
      It is not clear, what is the generic rule of transformation of the matrix elements of the evolution method
      -- one needs to study more complicated families of knots to even make a guess.
  }

  { %% ### When do transitions happen?
    {Transitions between regions, where evolution is valid, seem to happen when
      something ``essential'' happens to the planar diagram as a result of adding/removing a particular
      crossing.}

    {For instance, when we go from the point $(1,1)$ to the point $(2,1)$ the knot goes from two unknots
      which are not mutually linked to one unknot
      
      %% \begin{verbatim}
      %% The picture of these two cases of the knot diagram
      %% \end{verbatim}
      
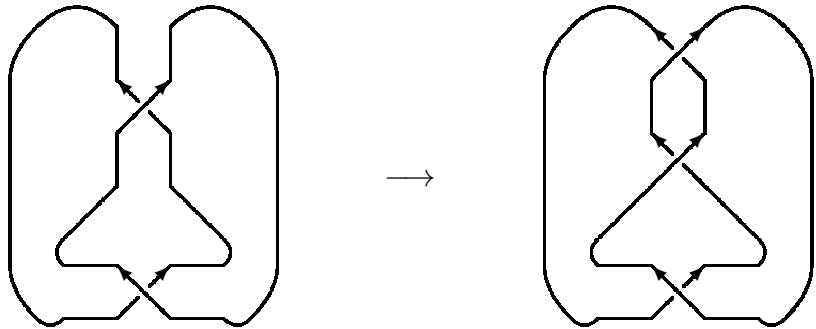
\begin{picture}(300,130)(-60,-5)
        \put(0,0){ %% ### A copy of 'vanilla' figure 8
          \thicklines
          \figureEightLikeTemplate
          \put(20,70){
            \crossingPosUp
            \put(0,-20){\line(0,1){20}}
            \put(0,20){\line(0,1){20}}
            \put(20,-20){\line(0,1){20}}
            \put(20,20){\line(0,1){20}}
          }
          \put(20,0){
            \crossingNegUp
            \put(-20,0){\line(1,0){20}}
            \put(20,0){\line(1,0){20}}
            \put(-20,20){\line(1,0){20}}
            \put(20,20){\line(1,0){20}}
          }
        }
        \put(120,50){$\longrightarrow$}
        \put(200,0){
          \thicklines
          \figureEightLikeTemplate
          \put(20,70){
            \put(0,-20){\crossingPosUp}
            \put(0,20){\crossingPosUp}
            \put(0,0){\line(0,1){20}}
            \put(20,0){\line(0,1){20}}
          }
          \put(20,0){
            \crossingNegUp
            \put(-20,0){\line(1,0){20}}
            \put(20,0){\line(1,0){20}}
            \put(-20,20){\line(1,0){20}}
            \put(20,20){\line(1,0){20}}
          }
        }
      \end{picture}
    }

    {\noindent and in transition from $(1,3)$ to $(0,3)$
      Hopf link goes into unknot

      %% \begin{verbatim}
      %% I'd better have a good picture of this
      %% \end{verbatim}
      
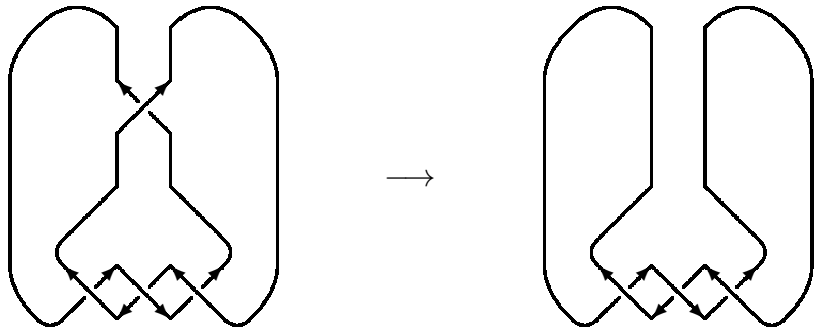
\begin{picture}(300,130)(-60,-5)
        \put(0,0){ %% ### A copy of 'vanilla' figure 8
          \thicklines
          \figureEightLikeTemplate
          \put(20,70){
            \crossingPosUp
            \put(0,-20){\line(0,1){20}}
            \put(0,20){\line(0,1){20}}
            \put(20,-20){\line(0,1){20}}
            \put(20,20){\line(0,1){20}}
          }
          \put(0,0){\crossingNegUp}
          \put(20,0){\crossingNegDown}
          \put(40,0){\crossingNegUp}
        }
        \put(120,50){$\longrightarrow$}
        \put(200,0){
          \thicklines
          \figureEightLikeTemplate
          \put(20,70){
            \put(0,-20){\line(0,1){60}}
            \put(20,-20){\line(0,1){60}}
          }
          \put(0,0){\crossingNegUp}
          \put(20,0){\crossingNegDown}
          \put(40,0){\crossingNegUp}
        }
      \end{picture}

      \noindent it is not at all clear where and when the phase transitions occur in general
      -- one needs to study more examples.
     }
  }

}

  {\section{Caveats in using KnotTheory package to calculate Khovanov polynomials}
    \label{sec:caveats}
    %% \fixme{actually write what the caveat is about}
    Even though the program ``Kh'' that is included in the ``KnotTheory'' package for Wolfram Mathematica
    calculates most Khovanov polynomials very fast and answers are correct,
    it somehow makes mistakes for the diagrams that contain a small number of crossings.

    For instance, it incorrectly calculates Khovanov polynomial of an unknot whose diagram contains just one crossing.
    Since we are no experts in Java, in which the program is actually written, and cannot debug it,
    we've used a workaround: we inserted an extra two-strand braid in our diagram, that contained alternating
    positive and negative crossings. Thanks to the second Reidemeister move such planar diagram is equivalent
    to the original diagram, so the Khovanov polynomial should not change. But it does have more crossings
    and the ``Kh'' program does not err on it.
  }

  {\section{Proof} \label{sec:proof}
    %% \fixme{No proof}
    In this section we prove the Theorem~\ref{thm:evolution-method}.

    First of all, note, that all knots and links in family that we consider are alternating.

    Second, for alternating knots and links, there is a theorem (\cite[Theorem 4.5]{Lee-endomorphism})
    that allows one to express Khovanov polynomial through the Jones one.
    For completeness, we reproduce it here.
    \begin{theorem}\cite{Lee-endomorphism} \label{for-thm:alternating-reconstruction}
      For an $n$ component oriented nonsplit alternating link $L$ with its components
      $S_1, \dots, S_n$ and linking numbers $l_{jk}$ of $S_j$ and $S_k$,
      \begin{align}
        Kh(L)(q,t) = q^{-\sigma(L)}
        \left ( (q + q^{-1})
        \left (\sum_{E \subset \{2,\dots n\}} (t q^2)^{\sum_{j \in E, k \notin E} 2 l_{jk}} \right)
        + \left ( q^{-1} + t q^2 \cdot q \right ) Kh'(L)(t q^2)
        \right )
      \end{align}
      for some polynomial $Kh'(L)$.
    \end{theorem}

    In our case link has at most two components.
    The signature, depending on the region of the parameter plane, equals
    \begin{align} \label{eq:signature-for-quadrants}
      \sigma_{UR} & = a - 2 \\ \notag
      \sigma_{UL} & = a \\ \notag
      \sigma_{LR} & = a \\ \notag
      \sigma_{LL} & = a + 2
    \end{align}

    For even $a$ our diagram is a knot, whereas for odd $a$ it is a two-component link,
    with the components' linking number $(b-a)/2$, so, for the sum over
    subsets~$E \subset \{2,\dots n\}$
    we can write
    \begin{align}
      1 + \frac{1}{2} \left ( 1 -(-1)^a \right) (t q^2)^{b-a}
    \end{align}

    So, for Khovanov and Jones polynomials in, say, UL region
    (one needs to adjust the value of the signature for the other regions)
    we can write
    \begin{align} \label{eq:khovanov-and-jones-reduced}
        Kh = q^{-a} \left ( (q + q^{-1})
        \left (1 + \frac{1}{2} \left ( 1 -(-1)^a \right) (t q^2)^{b-a} \right)
        + \left ( q^{-1} + t q^2 \cdot q \right ) Kh'(t q^2)
        \right )
        \\ \notag
        J = q^{-a} \left ( (q + q^{-1})
        \left (1 + \frac{1}{2} \left ( 1 -(-1)^a \right) (- q^2)^{b-a} \right)
        + \left ( q^{-1} - q^2 \cdot q \right ) J'(- q^2)
        \right ),
    \end{align}
    where polynomials $Kh'$ and $J'$ are actually equal, so if one considers
    $J'$ as a function of $q$ one can restore $Kh'$ by substituting $q \rightarrow q \sqrt{-t}$.
    
    Now, for the Jones polynomial we can obtain the following evolution
    on the whole parameter plane (using, for instance, RT-formalism)
    \begin{align}
      J_{\substack{a = 2l \hfill \\ b = 2k + 1}} & =
      \left(\begin{array}{cc} (1)^b & (- q^2)^b \end{array}\right)
      N
      \left(\begin{array}{c} q^{-a} \\ (- q^3)^{-a} \end{array}\right) \\ \notag
      N & =
      \left(
      \begin{array}{cc}
        \frac{\left(q^2 + 1 + q^{-2}\right)}{\left(q+q^{-1}\right)} &
        \frac{1}{q+q^{-1}} \\
        -\frac{\left(q^2 + 1 + q^{-2}\right)}{\left(q+q^{-1}\right)} &
        \frac{\left(q^2 + 1 + q^{-2}\right)}{\left(q+q^{-1}\right)}
        \\
      \end{array}
      \right)
    \end{align}

    Using formula \eqref{eq:khovanov-and-jones-reduced} to first express
    $J'$ through $J$, then $Kh'$ through $J'$ and finally obtain $Kh$
    through $a$ and $b$ one can straightforwardly check that, indeed,
    $Kh$ depends on $a$ and $b$ as in formulas \eqref{eq:evolution-formula} and \eqref{eq:evolution-matrices}.
    This completes the proof of the theorem.
  }

{\section{Conclusion}
  {In this paper we've built on the results of \cite{AnoM} by considering what
    happens to the evolution method for Khovanov polynomial for the simplest
    non-torus family of knots (see Section~\ref{sec:fig-eight-like-knots}),
    that in particular includes two-strand and twist knots.}

  {For this two-parametric family we found a peculiar chamber structure (see Section~\ref{sec:phase-diagram}),
    which generalizes ``mirror anomaly'' observed in \cite{AnoM}.
    We also found an unexpected third eigenvalue of the (hypothetical) fundamental $R$-matrix,
    which drops out in the limit $(t \rightarrow -1)$.}

  %% {Our evolution formulas are in agreement with BNGKh-conjecture (\cite[Conjecture 1]{BarNatan}),
  %%   which may serve as weak evidence in favor of both conjectures.
  %% }
  { At the heart of the proof of our evolution formulas is the theorem~\cite{\fixme{???}}
    that relates Khovanov polynomial for any alternating knot to its Jones polynomial.
    While computer experiments clearly show that evolution method extends beyond alternating knots,
    new ideas are required to extend the proof, even to the case of 2-strand braid insertion into \textit{arbitrary} knot.
  }
  
  {It is very interesting, what is the relevant generalization of the Reshetikhin-Turaev formalism.
    It should be capable of describing the observed chamber structure.
    The abrupt jumps of the evolution coefficients that occur when one goes between the chambers
    may hint that another deformation (in addition to $q$- and $t$-deformations) is required to
    smoothen these jumps out and embed the problem into the framework of the usual linear algebra.
    This indication is in accordance with recent developments of the representation theory of DIM algebra \cite{ZenDIM},
    where one more deformation parameter also begs to be introduced.}

  {At the moment it is not clear where exactly does the simple evolution break down.
    There is only a vague idea that it breaks when ``something interesting'' happens to the planar diagram
    -- either it goes from being disconnected to being connected, or it untwists in an unusual manner
    (see Section~\ref{sec:phase-diagram}).
  }

  {Study of more complicated families of knots is needed to clarify the situation.
    To perform such a study one needs better computer programs that allow to specify families of knots
    more conveniently
    (manually working out through all the odd-even cases for parameters and figuring orientations of strands is a bit tedious).
    We continue to work in this direction.}
}

{}

\end{document}